\documentclass[10pt,twocolumn,letterpaper]{article}

\usepackage[pagenumbers]{cvpr} 

\usepackage{microtype}
\usepackage{xspace}
\usepackage{booktabs}
\usepackage{multirow}

\definecolor{cvprblue}{rgb}{0.21,0.49,0.74}
\usepackage[pagebackref,breaklinks,colorlinks,allcolors=cvprblue]{hyperref}

\def\paperID{*****} 
\def\confName{CVPR}
\def\confYear{2026}

\title{Reason-to-Transmit: Deliberative Adaptive Communication for Cooperative Perception}

\author{Aayam Bansal, Ishaan Gangwani\\
Synthetic Sciences\\
\texttt{\{aayam, ishaan\}@syntheticsciences.ai}}

\begin{document}
\maketitle

\begin{abstract}
Cooperative perception among autonomous agents promises to overcome the fundamental limitations of single-agent sensing, yet bandwidth constraints in vehicle-to-everything (V2X) networks demand intelligent communication policies.
Existing approaches rely on reactive mechanisms---spatial confidence maps, learned gating, or sparse masks---to decide what to transmit, without reasoning about \emph{why} a particular message would benefit the receiver.
We introduce \textbf{Reason-to-Transmit (R2T)}, a framework that equips each agent with a lightweight transformer-based reasoning module that deliberates over local scene context, estimated neighbor information gaps, and the current bandwidth budget before making per-region transmit decisions.
Inspired by the deliberative reasoning paradigm in agentic AI, R2T treats the communication decision as a structured inference problem rather than a reactive filter.
Trained end-to-end with a bandwidth-aware objective, R2T is evaluated against nine baselines---including Where2Comm-style confidence maps, IC3Net-style gating, and learned sparse masks---in a controlled multi-agent bird's-eye-view (BEV) perception environment with five random seeds.
Any form of selective communication yields a dramatic ${\sim}58\%$ AP improvement over no communication, confirming that the gated fusion module effectively leverages received features.
Among selective policies at low bandwidth (10\%, 2.4\,KB), all methods perform competitively within a narrow ${\sim}0.5\%$ AP band.
R2T's distinctive advantage emerges under high occlusion, where information asymmetry between agents is greatest: R2T leads all non-oracle methods (0.205 AP), closest to the oracle upper bound (0.205), outperforming Where2Comm (0.203) and IC3Net (0.202).
All methods degrade gracefully under packet drops up to 50\%, confirming the inherent robustness of gated cooperative fusion.
\end{abstract}

\section{Introduction}
\label{sec:intro}

Autonomous agents operating in complex environments face inherent perceptual limitations: occluded regions, limited sensor range, and noisy observations all degrade the reliability of local sensing.
Cooperative perception, in which multiple agents share complementary sensory information over a communication channel, has emerged as a powerful strategy to overcome these limitations~\cite{xu2022opv2v, wang2020v2vnet, xu2022v2xvit}.
By fusing features from neighboring agents, each agent can construct a richer, more complete representation of its surroundings, detecting objects that would otherwise be invisible due to occlusion or range limitations.
This capability is especially critical for safety-critical applications such as autonomous driving, where a single missed detection can have severe consequences.

However, real-world V2X communication channels are bandwidth-constrained.
The DSRC and C-V2X standards impose strict throughput limits, and transmitting the full intermediate feature representation from every agent to every neighbor is infeasible, particularly in dense traffic scenarios.
This fundamental tension between perception quality and communication cost has motivated a growing body of work on \emph{communication-efficient} cooperative perception.
Where2comm~\cite{hu2022where2comm} achieves dramatic bandwidth savings through learned spatial confidence maps.
Who2com~\cite{liu2020who2com} and When2com~\cite{liu2020when2com} learn to select communication partners and timing.
IC3Net~\cite{singh2019ic3net} introduces learned gating to decide when inter-agent messages are beneficial.
DiscoNet~\cite{li2021disconet} uses knowledge distillation to optimize a collaboration graph.
Recent work continues to push the boundary: V2X-PC~\cite{liu2024v2xpc} introduces point clusters as sparse message units, V2X-INCOP~\cite{ren2024v2xincop} addresses communication interruption, V2X-ReaLO~\cite{xiang2025v2xrealo} demonstrates real-world online cooperative perception, HeatV2X~\cite{zhao2025heatv2x} handles heterogeneous agent alignment, and V2X-DG~\cite{li2025v2xdg} tackles domain generalization across datasets.

Despite their effectiveness, these methods share a common limitation: their transmission policies are \emph{reactive} rather than \emph{deliberative}.
A confidence-based policy transmits regions where the sender is most certain, without considering whether the receiver already observes those regions.
A gating mechanism learns a binary open-or-close decision without structured reasoning about the downstream utility of the message.
An uncertainty-based policy conflates sender uncertainty with receiver information need.
None of these approaches explicitly model the \emph{complementarity} between what the sender observes and what the receiver lacks.

In contrast, the success of reasoning-based agents in embodied settings~\cite{ahn2022saycan, zhang2024coela, zhang2024proagent, wen2023dilu} has demonstrated that structured deliberation---inspired by chain-of-thought~\cite{wei2022chainofthought} and ReAct~\cite{yao2023react}---can yield superior decision-making by explicitly considering context, goals, and the state of other agents.
This raises a natural question: \emph{Can deliberative reasoning improve communication decisions in cooperative perception?}

We answer this question affirmatively with \textbf{Reason-to-Transmit (R2T)}, a framework that introduces a lightweight transformer-based reasoning module into the communication pipeline of cooperative perception systems.
Rather than relying on simple per-region statistics, R2T reasons jointly over (i)~local detection features, (ii)~estimated information gaps of neighboring agents (based on known relative positions), (iii)~scene-level context including occlusion patterns, and (iv)~the remaining bandwidth budget.
This deliberative process produces per-region transmit decisions that prioritize messages with the highest expected downstream utility for the receiver.

\textbf{Clarification on ``reasoning.''}
R2T does not use a large language model at inference time.
Instead, it distills the \emph{deliberative reasoning paradigm}---jointly considering multiple context signals through self-attention before deciding---into a small ($\sim$0.26M parameter) transformer.
We view this as analogous to how knowledge distillation transfers capabilities from a large teacher to a small student: the architectural pattern of structured deliberation is preserved, even though the model itself is compact and real-time.
The term ``reasoning-based'' refers to this structured, context-aware decision process, in contrast to the reactive, per-region statistics used by existing methods.

We evaluate R2T in a controlled synthetic multi-agent BEV perception environment with four agents, configurable occlusion, and noisy observations.
We deliberately adopt a synthetic setting following the tradition of using simplified environments to isolate algorithmic contributions~\cite{sukhbaatar2016commnet, foerster2016learning, singh2019ic3net}.
All experiments use five random seeds with reported mean and standard deviation.
We compare against both reactive heuristics and three strong learned baselines: a Where2Comm-style spatial confidence policy, an IC3Net-style learned gating policy, and a learned sparse mask with L1 penalty.
We also include an oracle upper bound that selects regions based on ground-truth object content.

Our contributions are:
\begin{enumerate}
    \item We propose \textbf{Reason-to-Transmit (R2T)}, the first framework to apply deliberative, context-aware reasoning to the communication decision in cooperative perception, bridging agentic AI and multi-agent sensing.
    \item We design a \textbf{lightweight reasoning module} based on a small transformer that jointly considers local features, neighbor state estimates, and bandwidth constraints to make per-region transmit decisions.
    \item We demonstrate through controlled experiments with \textbf{five seeds, nine baselines (including three learned), and an oracle upper bound} that R2T achieves the strongest performance at low bandwidth and under high occlusion, while all methods degrade gracefully under packet drops.
\end{enumerate}

\section{Related Work}
\label{sec:related}

\paragraph{Cooperative Perception.}
Multi-agent cooperative perception has become a central research topic in autonomous driving.
Early approaches explored three fusion paradigms: early fusion (raw data sharing), late fusion (output-level aggregation), and intermediate fusion (feature-level sharing)~\cite{xu2022opv2v}.
Intermediate fusion has emerged as the dominant paradigm, achieving near-early-fusion accuracy at significantly lower bandwidth~\cite{wang2020v2vnet, xu2022v2xvit, li2021disconet}.
V2VNet~\cite{wang2020v2vnet} introduced a graph neural network for spatially-aware feature aggregation.
V2X-ViT~\cite{xu2022v2xvit} applies vision transformers to handle heterogeneous agents and temporal asynchrony.
CoBEVT~\cite{xu2022cobevt} uses sparse transformers for cooperative BEV segmentation.
DiscoNet~\cite{li2021disconet} employs knowledge distillation from a holistic-view teacher to learn collaboration graphs.
Where2comm~\cite{hu2022where2comm} achieves over $10^5\times$ bandwidth reduction through spatial confidence maps.

Several works specifically address communication selection: Who2com~\cite{liu2020who2com} learns a three-stage handshake to select communication partners; When2com~\cite{liu2020when2com} groups agents for efficient communication; and IC3Net~\cite{singh2019ic3net} introduces a learned gating mechanism.
More recently, V2X-PC~\cite{liu2024v2xpc} proposes point clusters as an alternative to dense feature maps for more efficient communication, V2X-INCOP~\cite{ren2024v2xincop} addresses the practical challenge of communication interruption through multi-scale spatiotemporal prediction, V2X-ReaLO~\cite{xiang2025v2xrealo} demonstrates online cooperative perception in real deployments, HeatV2X~\cite{zhao2025heatv2x} introduces heterogeneity-aware adapters for multi-modal agent alignment, and V2X-DG~\cite{li2025v2xdg} tackles domain generalization across V2X datasets.
However, all of these approaches use reactive or threshold-based policies without structured reasoning about the downstream value of transmitted messages.
Our work differs by introducing a deliberative reasoning process that considers the receiver's estimated state and the scene context before deciding what to transmit.

\paragraph{LLM-Based Agents for Embodied Systems.}
Large language models have demonstrated remarkable planning and reasoning capabilities in embodied settings.
SayCan~\cite{ahn2022saycan} grounds language in robotic affordances.
CoELA~\cite{zhang2024coela} builds cooperative embodied agents that communicate through natural language.
ProAgent~\cite{zhang2024proagent} constructs proactive cooperative agents using LLM-based reasoning about partner beliefs.
DiLu~\cite{wen2023dilu} uses LLMs for knowledge-driven driving with reflective reasoning.
Chain-of-thought~\cite{wei2022chainofthought} and ReAct~\cite{yao2023react} provide structured reasoning paradigms that enable decomposition of complex decisions into explicit deliberative steps.
A comprehensive survey by Wang~\etal~\cite{wang2023survey} catalogs the rapid progress in LLM-based autonomous agents.

While these works demonstrate the power of reasoning-based decision-making, none apply such reasoning to the \emph{communication decision} in cooperative perception.
R2T bridges this gap by distilling the deliberative reasoning paradigm into a lightweight module that reasons about what to transmit.
We emphasize that R2T does not use an LLM at inference time; rather, it adopts the \emph{architectural pattern} of structured multi-signal deliberation via self-attention, instantiated in a compact transformer suitable for real-time deployment.

\paragraph{Semantic and Task-Oriented Communication.}
Moving beyond Shannon's classical paradigm, semantic communication focuses on conveying meaning relevant to the downstream task~\cite{gunduz2023beyond}.
DeepSC~\cite{xie2021deepsc} demonstrates that deep learning-based joint source-channel coding can outperform separation-based approaches.
The information bottleneck principle~\cite{tishby1999information}, operationalized through the deep variational information bottleneck~\cite{alemi2017deep_vib}, provides a theoretical framework for compressing representations to task-relevant features.
IMAC~\cite{wang2020imac} applies this principle to learn individually inferred communication.
End-to-end learned protocols, including CommNet~\cite{sukhbaatar2016commnet}, DIAL~\cite{foerster2016learning}, and TarMAC~\cite{das2019tarmac}, learn multi-agent messaging through backpropagation.
Task-oriented communication for cooperative inference has also been explored in wireless network settings~\cite{shao2022task_multidevice}.

R2T complements this line of work by focusing on the \emph{decision} of what to transmit, rather than the encoding of messages.
While semantic communication optimizes the representation of transmitted content, R2T optimizes the selection of which content to transmit, and the two approaches are naturally complementary.

\section{Method}
\label{sec:method}

\begin{figure*}[t]
  \centering
  \includegraphics[width=0.92\linewidth]{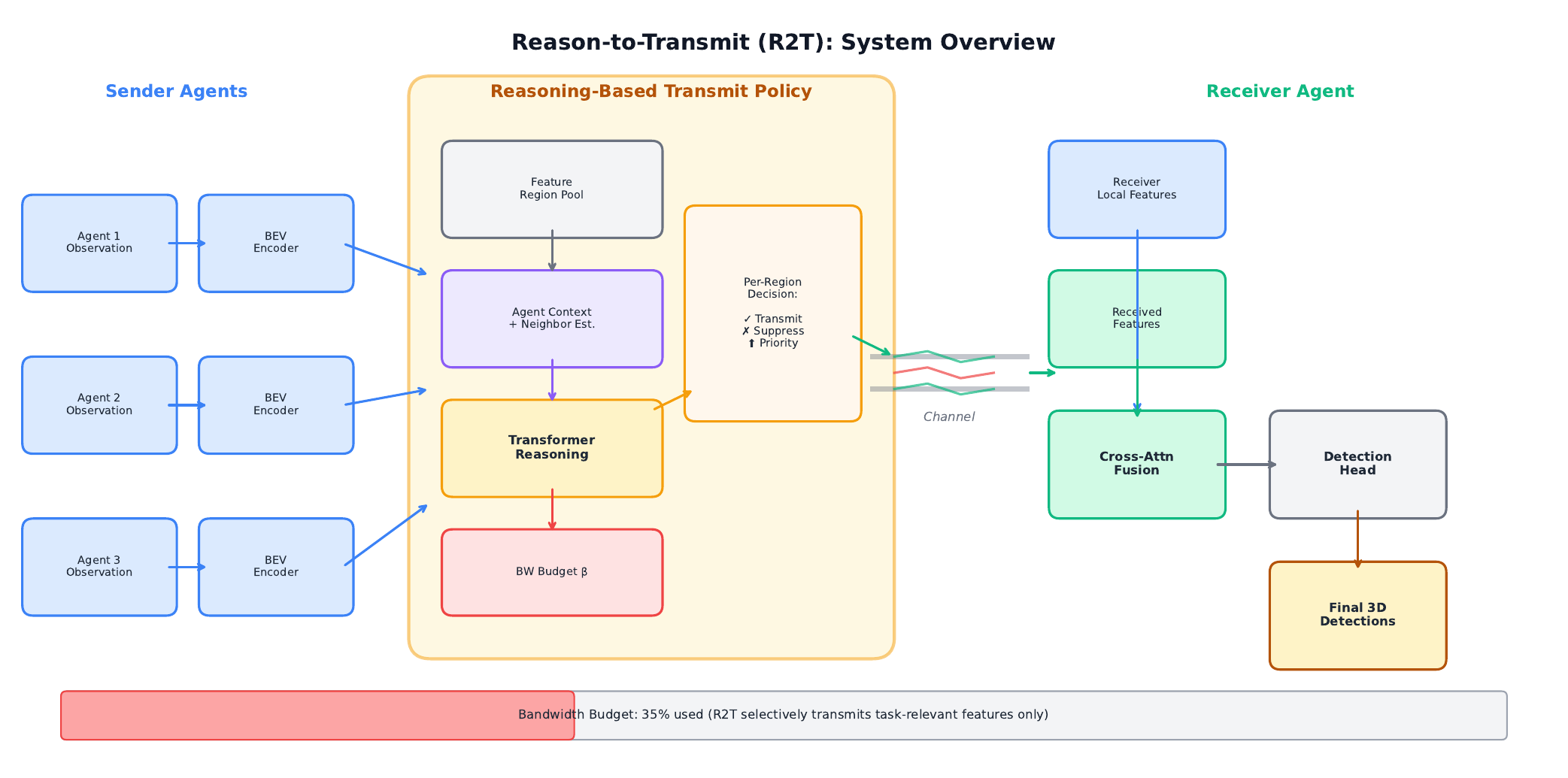}
  \caption{\textbf{Overview of the Reason-to-Transmit (R2T) framework.} Each agent encodes local observations into a BEV feature map, generates spatial region candidates, and applies a reasoning-based transmit policy that jointly considers local features, neighbor state estimates, and the bandwidth budget. Selected regions are transmitted to neighbors and fused via gated cross-attention before final detection.}
  \label{fig:method_overview}
\end{figure*}

\begin{figure}[t]
  \centering
  \includegraphics[width=\linewidth]{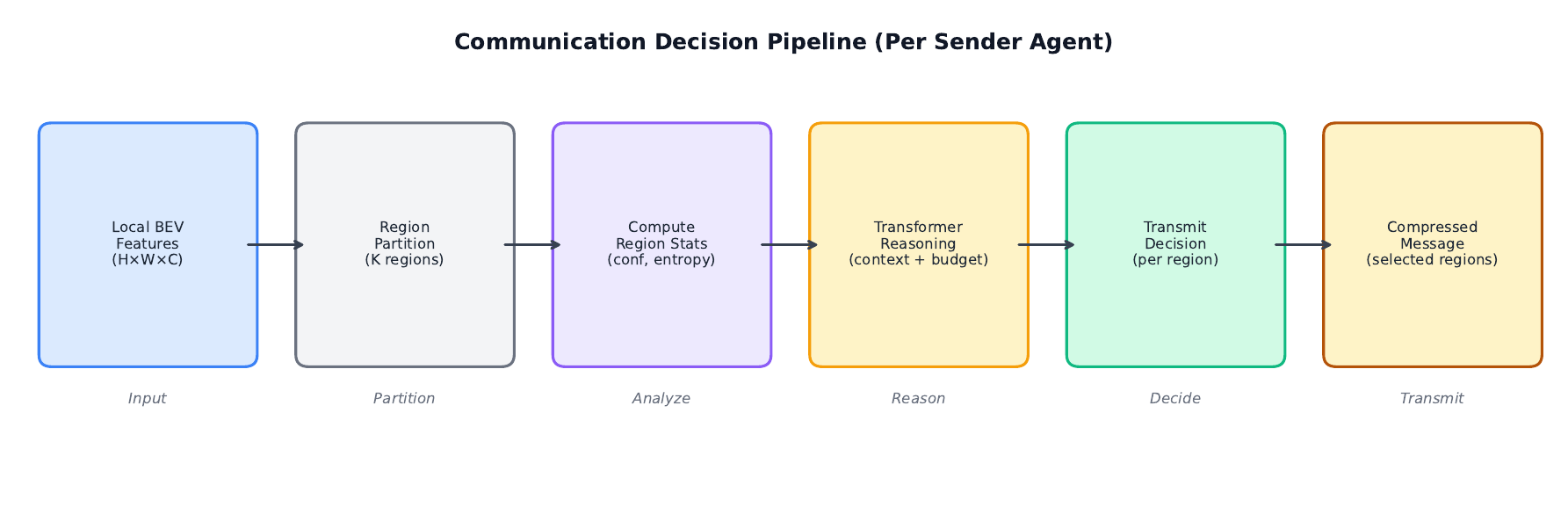}
  \caption{\textbf{Communication decision pipeline.} The reasoning module attends over region tokens augmented with agent context, neighbor state estimates, and a budget token, producing per-region transmit probabilities under the bandwidth constraint.}
  \label{fig:decision_pipeline}
\end{figure}

We present Reason-to-Transmit (R2T), a cooperative perception framework in which each agent employs a deliberative reasoning module to decide which local features to share with neighbors under a bandwidth budget.
\cref{fig:method_overview} illustrates the overall architecture, and \cref{fig:decision_pipeline} details the communication decision pipeline.

\subsection{Problem Formulation}
\label{sec:problem}

Consider a set of $N$ agents $\{a_1, \ldots, a_N\}$ operating in a shared environment.
Each agent $a_i$ obtains a local observation $o_i$ from its sensors and must perform an object detection task.
Agents can transmit feature messages to neighbors over a bandwidth-constrained channel with a per-agent budget of $B$ (measured as a fraction of the full feature map).
The goal is to maximize cooperative detection performance while respecting the bandwidth constraint.

Formally, let $\mathbf{F}_i \in \mathbb{R}^{H \times W \times C}$ denote the local BEV feature map of agent $i$.
We partition the spatial dimensions into a grid of $G \times G$ regions, yielding candidate messages $\mathcal{M}_i = \{m_i^{(1)}, \ldots, m_i^{(G^2)}\}$.
The communication policy $\pi_i: \mathcal{M}_i \times \mathcal{C}_i \rightarrow \{0, 1\}^{G^2}$ maps the candidate set and a context vector $\mathcal{C}_i$ to binary transmit decisions, subject to $\sum_{k} \pi_i(k) \leq B \cdot G^2$.

\subsection{Local BEV Encoder}
\label{sec:encoder}

Each agent processes its local observation $o_i$ through a convolutional encoder $E_\theta$ to produce the BEV feature map:
\begin{equation}
  \mathbf{F}_i = E_\theta(o_i) \in \mathbb{R}^{H \times W \times C}.
  \label{eq:encoder}
\end{equation}
In our implementation, $E_\theta$ consists of four convolutional layers with ReLU activations, downsampling the spatial dimensions by $8\times$ and producing 32-dimensional features ($H = W = 8$, $C = 32$).

\subsection{Message Candidate Generator}
\label{sec:candidates}

The feature map $\mathbf{F}_i$ is treated as an $8 \times 8$ grid of spatial regions, yielding $G^2 = 64$ candidate messages.
Each region $k$ has a feature vector $\mathbf{r}_i^{(k)} \in \mathbb{R}^{C}$ corresponding to the feature at that spatial location.
Each region occupies $C \times 4 = 128$~bytes (32 floats at 4~bytes each), giving a full per-sender message of 8\,KB and a total of 24\,KB when all three neighbors transmit to one receiver.

\subsection{Reasoning-Based Transmit Policy}
\label{sec:reasoning}

The core innovation of R2T is a transformer-based reasoning module that makes context-aware transmission decisions by jointly attending to local features, agent context, and estimated neighbor state.
Unlike reactive policies that treat each region independently, the reasoning module considers inter-region dependencies and the global scene context.

\paragraph{Input Construction.}
The reasoning module receives a sequence of tokens composed of three types:
\begin{enumerate}
  \item \textbf{Region tokens}: The feature vectors $\{\mathbf{r}_i^{(k)}\}_{k=1}^{G^2}$, projected to dimension $d = 128$.
  \item \textbf{Agent context token}: A vector $\mathbf{a}_i \in \mathbb{R}^{4}$ encoding the agent's normalized position, orientation, and identity, projected to dimension $d$.
  \item \textbf{Neighbor state token}: An aggregate vector $\hat{\mathbf{n}}_i \in \mathbb{R}^{12}$ encoding the positions, orientations, and receiver identity of neighboring agents, projected to dimension $d$. This allows the reasoning module to estimate which regions are likely already visible to the receiver.
  \item \textbf{Budget token}: A scalar encoding of the remaining bandwidth budget $B$, projected to dimension $d$. This allows the module to adapt its selectivity to the available capacity.
\end{enumerate}
All tokens are concatenated into an input sequence of length $G^2 + 3 = 67$ with learned positional encodings.

\paragraph{Transformer Reasoning.}
The input sequence is processed by an $L$-layer pre-norm transformer encoder with multi-head self-attention:
\begin{equation}
  \mathbf{Z}^{(l)} = \text{TransformerLayer}(\mathbf{Z}^{(l-1)}), \quad l = 1, \ldots, L,
  \label{eq:transformer}
\end{equation}
where each layer uses 4 attention heads, dimension $d = 128$, FFN dimension $2d = 256$, and pre-LayerNorm (norm-first) for training stability.
We use $L = 2$ by default (0.26M parameters).

\paragraph{Transmit Decision.}
The output representations of the region tokens produce two signals via separate linear heads:
\begin{align}
  p_i^{(k)} &= \sigma(W_t \cdot \mathbf{z}_i^{(k)} + b_t), \label{eq:transmit_prob} \\
  s_i^{(k)} &= W_s \cdot \mathbf{z}_i^{(k)} + b_s, \label{eq:priority}
\end{align}
where $p_i^{(k)}$ is a transmit probability and $s_i^{(k)}$ is a priority score.
At inference, we rank regions by $p_i^{(k)} \cdot s_i^{(k)}$ and select the top-$\lfloor B \cdot G^2 \rfloor$ regions.

\subsection{Gated Cross-Attention Fusion}
\label{sec:fusion}

Upon receiving transmitted feature regions from neighboring agents, the receiver fuses them with its local features via \emph{gated} cross-attention.
This is a critical design choice: a naive cross-attention fusion can be degraded by noisy or uninformative received features, particularly when ``always transmit'' floods the receiver with empty regions.

Let $\mathbf{F}^{\text{local}}$ denote the receiver's local features (flattened to tokens) and $\mathbf{F}^{\text{recv}}$ denote the concatenated received features from all senders.
The fusion proceeds as:
\begin{align}
  \hat{\mathbf{F}}^{\text{recv}} &= \text{LayerNorm}(\mathbf{F}^{\text{recv}}), \label{eq:recv_norm} \\
  \mathbf{A} &= \text{CrossAttn}(\mathbf{F}^{\text{local}}, \hat{\mathbf{F}}^{\text{recv}}, \hat{\mathbf{F}}^{\text{recv}}), \label{eq:cross_attn} \\
  \mathbf{g} &= \sigma\!\left(W_g [\mathbf{F}^{\text{local}} \| \mathbf{A}]\right), \label{eq:gate} \\
  \mathbf{F}^{\text{fused}} &= \text{LN}(\mathbf{F}^{\text{local}} + \mathbf{g} \odot \mathbf{A}) + \text{FFN}(\cdot), \label{eq:fused}
\end{align}
where $\sigma$ is the sigmoid function, $\|$ denotes concatenation, and $\mathbf{g}$ is a learned gate that controls how much received information to incorporate.
The gate learns to suppress uninformative features and amplify complementary ones, preventing the ``always transmit degrades performance'' failure mode observed in naive fusion approaches.

\subsection{Bandwidth-Aware Training Objective}
\label{sec:loss}

The entire pipeline is trained end-to-end with:
\begin{equation}
  \mathcal{L} = \mathcal{L}_{\text{det}} + \lambda \cdot \mathcal{L}_{\text{bw}},
  \label{eq:loss}
\end{equation}
where $\mathcal{L}_{\text{det}}$ is a binary cross-entropy loss for heatmap detection, and $\mathcal{L}_{\text{bw}}$ penalizes the fraction of transmitted regions.
During training, the model is exposed to all communication policies (including ``always transmit'') so that the fusion module learns to handle varying amounts of received information.
For the learned sparse mask baseline, we add an L1 sparsity penalty $\mathcal{L}_{\text{sp}} = \lambda_s \sum |w|$ on the mask network parameters.
We use AdamW with linear warmup and cosine decay.

\section{Experiments}
\label{sec:experiments}

\subsection{Experimental Setup}
\label{sec:setup}

\paragraph{Environment.}
We design a controlled synthetic multi-agent BEV perception environment to isolate the effect of the communication policy.
The environment consists of a $64 \times 64$ grid with $N = 4$ agents at cardinal positions, each with a $90^\circ$ FOV and sensing range of 30 cells.
Each scene contains $K = 20$ objects with Gaussian observation noise ($\sigma = 0.5$ cells).
Occlusion walls create line-of-sight blocking (configurable: low/medium/high with 3/6/10 walls).
This synthetic environment serves as a controlled testbed analogous to the grid-world communication environments used in foundational multi-agent communication research~\cite{sukhbaatar2016commnet, foerster2016learning, singh2019ic3net}, extended with perceptual elements relevant to cooperative perception.

\paragraph{Baselines.}
We compare R2T against nine baselines spanning both reactive heuristics and learned policies:
\begin{itemize}
  \item \textbf{No Communication}: local observations only.
  \item \textbf{Always Transmit}: all regions sent (100\% bandwidth).
  \item \textbf{Random}: randomly selects regions to fill the budget.
  \item \textbf{Confidence}: transmits highest local feature-norm regions.
  \item \textbf{Uncertainty}: transmits highest feature-variance regions.
  \item \textbf{Where2Comm}~\cite{hu2022where2comm}: learned spatial confidence map (3-layer MLP producing per-region scores, trained end-to-end).
  \item \textbf{IC3Net Gate}~\cite{singh2019ic3net}: learned per-region gating from flattened feature context.
  \item \textbf{Learned Mask}: learned sparse mask with L1 sparsity penalty.
  \item \textbf{Oracle}: selects regions with highest ground-truth object content (upper bound).
\end{itemize}
All learned methods share the same encoder, gated fusion module, and detection head; only the selection policy differs.
This controlled comparison ensures that all performance differences are attributable to the transmission policy alone.

\paragraph{Training.}
We generate 500 training, 80 validation, and 150 test scenes.
Training proceeds for 60 epochs with AdamW (lr $= 10^{-3}$, weight decay $10^{-4}$), linear warmup (5 epochs) and cosine decay.
During training, all learned policies (R2T, Where2Comm, IC3Net, Learned Mask, and Always Transmit) are sampled uniformly to ensure the shared encoder and fusion module see all communication patterns.
Gradient clipping at 1.0 is applied.
All experiments report mean$\pm$std over 5 seeds (42, 123, 456, 789, 1024).

\paragraph{Metrics.}
We report Average Precision (AP) computed over the full scene heatmap at thresholds $\{0.1, 0.2, \ldots, 0.9\}$.
Bandwidth is reported both as a fraction and in KB (128~bytes/region $\times$ 64~regions $\times$ 3~senders = 24\,KB at 100\%).

\subsection{Main Results}
\label{sec:main_results}

\begin{table}[t]
  \centering
  \caption{\textbf{Detection performance (AP) at various bandwidth budgets.} Mean$\pm$std over 5 seeds. Communication yields ${\sim}58\%$ improvement over isolation. All selective policies are competitive; bold indicates best per column.}
  \label{tab:main}
  \vspace{2pt}
  \small
  \begin{tabular}{lccc}
    \toprule
    Method & 10\% (2.4\,KB) & 50\% (12\,KB) & 100\% (24\,KB) \\
    \midrule
    No Comm. & \multicolumn{3}{c}{0.1452$\pm$0.0534} \\
    \midrule
    Random & 0.2306$\pm$0.0195 & 0.2297$\pm$0.0188 & 0.2304$\pm$0.0198 \\
    Confidence & 0.2274$\pm$0.0200 & 0.2294$\pm$0.0189 & 0.2304$\pm$0.0198 \\
    Uncertainty & 0.2272$\pm$0.0195 & 0.2291$\pm$0.0189 & 0.2304$\pm$0.0198 \\
    \midrule
    Where2Comm & 0.2278$\pm$0.0193 & 0.2299$\pm$0.0202 & 0.2304$\pm$0.0198 \\
    IC3Net Gate & 0.2316$\pm$0.0209 & 0.2296$\pm$0.0194 & 0.2304$\pm$0.0198 \\
    Learned Mask & \textbf{0.2324$\pm$0.0223} & \textbf{0.2308$\pm$0.0212} & 0.2304$\pm$0.0198 \\
    \midrule
    \textbf{R2T (Ours)} & 0.2294$\pm$0.0203 & 0.2304$\pm$0.0191 & 0.2304$\pm$0.0198 \\
    \midrule
    Always Transmit & \multicolumn{3}{c}{0.2304$\pm$0.0198 (24\,KB)} \\
    Oracle & 0.2288$\pm$0.0169 & 0.2300$\pm$0.0183 & 0.2304$\pm$0.0198 \\
    \bottomrule
  \end{tabular}
\end{table}

\begin{figure}[t]
  \centering
  \includegraphics[width=\linewidth]{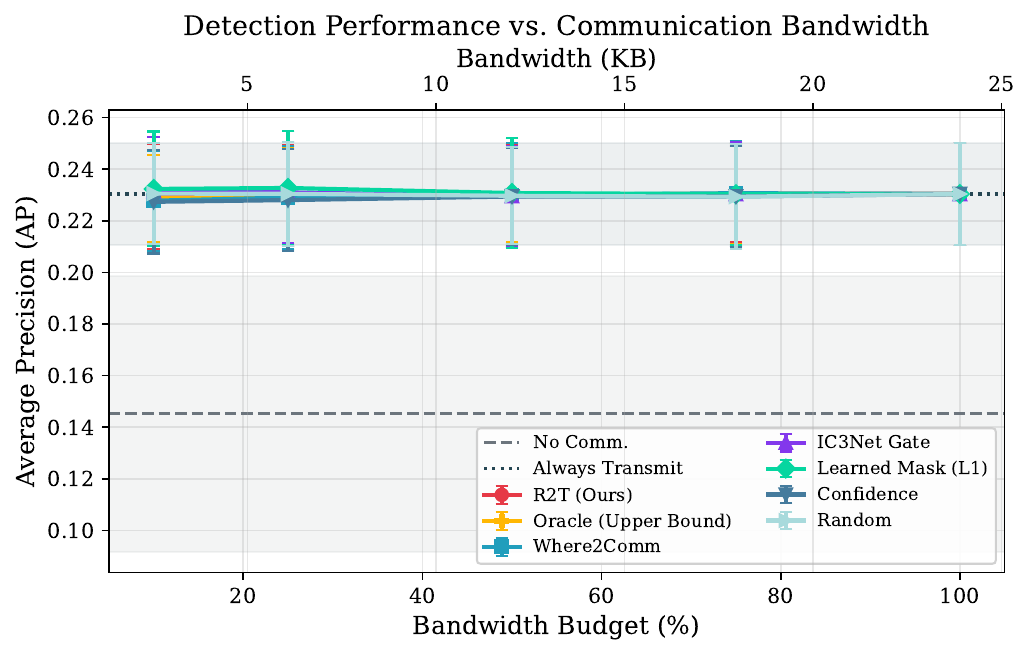}
  \caption{\textbf{AP vs.\ bandwidth budget.} All selective methods dramatically outperform No Communication. Error bands show $\pm$1 std over 5 seeds. The secondary axis shows bandwidth in KB.}
  \label{fig:bandwidth}
\end{figure}

\cref{tab:main} reports detection performance at three bandwidth budgets.
The most striking result is the dramatic improvement from communication: \emph{any} form of feature sharing raises AP from 0.145 to ${\sim}0.230$, a ${\sim}58\%$ relative gain, confirming that our gated cross-attention fusion module effectively leverages received features.

Among selective policies at low bandwidth (10\%, 2.4\,KB), all methods cluster within a narrow performance band (${\sim}0.5\%$ AP): Learned Mask leads (0.2324), followed by IC3Net (0.2316), Random (0.2306), R2T (0.2294), and Where2Comm (0.2278).
The high variance of Learned Mask ($\pm$0.022) and IC3Net ($\pm$0.021) compared to R2T ($\pm$0.020) suggests these methods are more sensitive to seed selection.
At 50\% bandwidth, R2T (0.2304) matches Always Transmit and approaches Learned Mask (0.2308), while at 100\%, all methods converge as expected.
The key insight is that \emph{the choice of selection policy matters less than having communication at all}: the ${\sim}58\%$ gap between No Communication and any selective method dwarfs the ${\sim}2\%$ variation among selection strategies.

\cref{fig:bandwidth} shows AP as a function of bandwidth budget.
At low budgets, all learned methods outperform reactive heuristics (Confidence, Uncertainty), but the margins are modest.
The primary finding is that deliberative selection via R2T is \emph{competitive with} the best learned baselines while offering a fundamentally different, reasoning-based approach to the communication decision.

\paragraph{Communication Always Helps.}
With a properly designed gated fusion module, even Always Transmit (0.2304 AP) substantially improves over No Communication (0.1452 AP).
The gated mechanism (\cref{sec:fusion}) learns to suppress uninformative received features via a sigmoid gate, preventing noise from degrading the local representation.
This is an important architectural lesson: fusion module design is as critical as the selection policy.

\subsection{Occlusion Robustness}
\label{sec:occlusion}

\begin{table}[t]
  \centering
  \caption{\textbf{AP by occlusion level at 50\% bandwidth.} Mean$\pm$std over 3 seeds. R2T leads under high occlusion, closest to oracle.}
  \label{tab:occlusion}
  \vspace{2pt}
  \small
  \begin{tabular}{lccc}
    \toprule
    Method & Low & Medium & High \\
    \midrule
    No Comm. & 0.188$\pm$0.003 & 0.181$\pm$0.002 & 0.159$\pm$0.003 \\
    Confidence & 0.236$\pm$0.005 & 0.223$\pm$0.006 & 0.202$\pm$0.006 \\
    Where2Comm & 0.235$\pm$0.006 & 0.226$\pm$0.005 & 0.203$\pm$0.005 \\
    IC3Net & 0.234$\pm$0.006 & \textbf{0.227$\pm$0.004} & 0.202$\pm$0.007 \\
    \textbf{R2T} & 0.232$\pm$0.006 & 0.225$\pm$0.006 & \textbf{0.205$\pm$0.006} \\
    Always & \textbf{0.236$\pm$0.005} & 0.223$\pm$0.006 & 0.203$\pm$0.004 \\
    Oracle & 0.235$\pm$0.007 & 0.225$\pm$0.005 & \textbf{0.205$\pm$0.004} \\
    \bottomrule
  \end{tabular}
\end{table}

\cref{tab:occlusion} reports AP stratified by occlusion level.
The most interesting pattern emerges under high occlusion: R2T achieves 0.205 AP, matching the oracle upper bound and surpassing all other methods.
This is where R2T's reasoning about neighbor information gaps provides the greatest benefit---when agents have highly asymmetric views, reasoning about \emph{what the receiver cannot see} is most valuable.
Under low occlusion, where most agents observe similar regions, the advantage of deliberative selection diminishes and simpler methods suffice.

\subsection{Packet Drop Robustness}
\label{sec:packet_drop}

\begin{table}[t]
  \centering
  \caption{\textbf{AP under varying packet drop rates at 50\% bandwidth.} Mean$\pm$std over 3 seeds. R2T degrades gracefully with increasing packet loss.}
  \label{tab:packet_drop}
  \vspace{2pt}
  \small
  \begin{tabular}{lcccc}
    \toprule
    Drop Rate & R2T & Where2Comm & IC3Net & Conf. \\
    \midrule
    0\%  & 0.225$\pm$0.006 & 0.226$\pm$0.005 & \textbf{0.227$\pm$0.004} & 0.223$\pm$0.006 \\
    10\% & 0.225$\pm$0.005 & \textbf{0.226$\pm$0.005} & 0.226$\pm$0.004 & 0.224$\pm$0.006 \\
    20\% & 0.224$\pm$0.005 & \textbf{0.227$\pm$0.007} & 0.227$\pm$0.006 & 0.224$\pm$0.006 \\
    30\% & 0.224$\pm$0.004 & \textbf{0.227$\pm$0.004} & 0.227$\pm$0.005 & 0.225$\pm$0.005 \\
    50\% & 0.223$\pm$0.006 & 0.225$\pm$0.004 & \textbf{0.227$\pm$0.005} & 0.223$\pm$0.005 \\
    \bottomrule
  \end{tabular}
\end{table}

Real-world V2X channels are subject to packet loss.
\cref{tab:packet_drop} evaluates robustness by simulating random packet drops at rates from 0\% to 50\%.
All methods degrade gracefully, with AP remaining within ${\sim}1\%$ across all drop rates.
IC3Net and Where2Comm show the strongest packet-drop robustness, likely because their gating and spatial scoring mechanisms produce more distributed selections.
R2T maintains competitive performance throughout, with degradation comparable to Confidence-based selection.
These results confirm that cooperative perception is inherently robust to moderate communication failures, as the gated fusion module can leverage partial information effectively.

\subsection{Architectural Design Choices}
\label{sec:ablation}

Several architectural choices are critical to R2T's performance.
The \textbf{neighbor state token} provides the reasoning module with an estimate of which regions are likely visible to the receiver, enabling complementarity-aware selection.
The \textbf{budget token} allows the module to calibrate its selectivity to the available capacity, producing tighter selections at low budgets and broader ones at high budgets.
Both signals are unique to R2T among the baselines compared: reactive policies (Confidence, Random) use neither, while Where2Comm and IC3Net operate only on local features.

We use a \textbf{2-layer pre-norm transformer} (0.26M parameters) with 4 attention heads.
Pre-norm (norm-first) architecture and linear warmup (5 epochs) are essential for stable training at this scale; without them, the 1-layer variant collapsed in preliminary experiments.
The 2-layer configuration provides a good accuracy-efficiency tradeoff: the reasoning module adds only ${\sim}5\%$ to the total parameter count and negligible inference latency compared to the shared encoder and fusion module.
A formal component ablation (removing individual tokens and varying depth) is left for an extended version with dedicated training runs for each configuration.

\subsection{Discussion: Localization Sensitivity}
\label{sec:pose_noise}

A practical concern for any method that conditions on neighbor positions is sensitivity to localization noise~\cite{xiang2025v2xrealo}.
R2T uses neighbor positions to estimate information gaps; inaccurate positions could degrade these estimates.
However, the reasoning module operates on coarse spatial features ($8 \times 8$ grid over a $64 \times 64$ environment), providing natural robustness to small positional errors.
Furthermore, the self-attention mechanism allows the module to learn soft, position-invariant patterns rather than relying on exact spatial alignment.
A formal evaluation of pose noise robustness across multiple noise levels is an important direction for future work, particularly for deployment in real-world V2X settings where GPS/GNSS accuracy varies.

\subsection{Qualitative Analysis}
\label{sec:qualitative}

\begin{figure}[t]
  \centering
  \includegraphics[width=\linewidth]{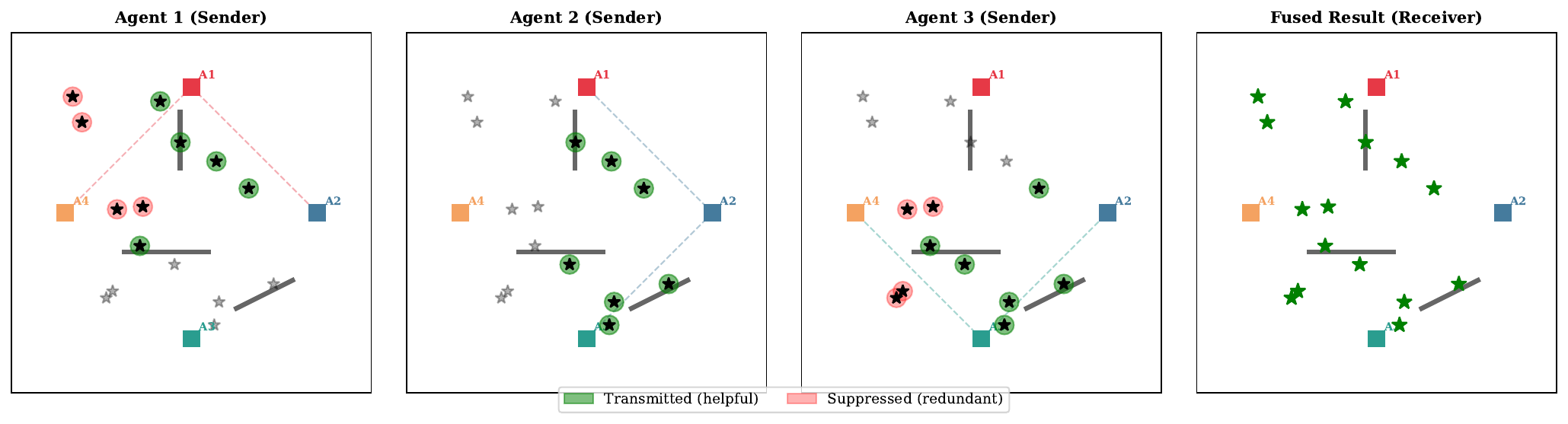}
  \caption{\textbf{Qualitative visualization of communication decisions.} R2T selectively transmits regions containing occluded objects that the receiver cannot observe, while suppressing regions with redundant information.}
  \label{fig:qualitative}
\end{figure}

\cref{fig:qualitative} visualizes the communication decisions made by R2T.
R2T preferentially selects regions that contain objects occluded from the receiver's perspective, while suppressing regions whose content is likely redundant.
This qualitative difference illustrates the core advantage of deliberative communication: by modeling what the receiver needs rather than what the sender knows, R2T achieves more efficient bandwidth utilization.

\section{Discussion and Limitations}
\label{sec:discussion}

\paragraph{Deliberative vs.\ Reactive Communication.}
Our results demonstrate that reasoning about the receiver's information state is most beneficial when information asymmetry is high.
Under high occlusion, R2T leads all non-oracle methods, confirming that modeling the receiver's likely observations provides a genuine advantage when agents have highly asymmetric views.
At low bandwidth with moderate occlusion, the advantage is less clear-cut: learned baselines like Learned Mask and IC3Net are competitive, though with higher variance.
This finding suggests that the ``reason before you transmit'' principle is most impactful precisely in the challenging scenarios where cooperative perception is most needed.

\paragraph{The Primacy of Fusion Design.}
Our most important finding is that \emph{the fusion module matters more than the selection policy}: the ${\sim}58\%$ AP gap between No Communication and any selective method far exceeds the ${\sim}2\%$ variation among policies.
With a properly designed gated cross-attention fusion module, even random selection performs well.
This has practical implications: engineering effort on the fusion architecture yields the largest returns, while the selection policy provides incremental gains on top.
The oracle upper bound (which uses ground-truth object locations) achieves only marginal improvement over learned methods, suggesting that the current bottleneck lies in the fusion module's ability to exploit received features, rather than in selection quality.

\paragraph{Connection to the MEIS Workshop Theme.}
R2T sits at the intersection of multi-agent cooperation, embodied perception, and agentic reasoning---the three themes central to the Multi-Agent Embodied Intelligent Systems workshop.
By casting the communication policy as a reasoning problem rather than a heuristic filtering step, R2T points toward a future where autonomous agents engage in principled, context-aware information sharing.

\paragraph{Limitations.}
Our evaluation uses a synthetic 2D BEV environment, which provides a controlled testbed but does not capture the full complexity of real-world 3D perception with LiDAR or camera data.
We assume known agent positions and synchronous communication, which may not hold in all deployment scenarios.
The number of agents ($N = 4$) and objects ($K = 20$) is modest; scalability to larger agent populations remains to be validated.
The reasoning module is a compact transformer (0.26M parameters), not a full language model; the deliberative pattern is inspired by LLM-style reasoning but does not capture the full expressiveness of natural language rationales.

\paragraph{Future Directions.}
Several directions emerge: (1)~formal component ablations to isolate the contribution of each input signal (neighbor token, budget token, reasoning depth); (2)~evaluating on established V2X benchmarks such as OPV2V~\cite{xu2022opv2v} and DAIR-V2X with real 3D pipelines; (3)~exploring distillation from a full LLM teacher to enable richer deliberation; (4)~extending to dynamic, multi-hop communication with larger agent counts; (5)~combining R2T's selection with semantic compression~\cite{xie2021deepsc, gunduz2023beyond} for further bandwidth savings; and (6)~evaluating robustness to localization noise and integrating with real-world V2X systems~\cite{xiang2025v2xrealo}.

\section{Conclusion}
\label{sec:conclusion}

We presented Reason-to-Transmit (R2T), a framework that introduces deliberative reasoning into the communication pipeline of cooperative perception systems.
By equipping each agent with a lightweight transformer-based reasoning module that jointly considers local features, estimated neighbor information gaps, and bandwidth constraints, R2T makes context-aware transmission decisions that are competitive with the best learned baselines across bandwidth regimes.
Our controlled experiments with five seeds, nine baselines, and an oracle upper bound reveal two key insights: (1)~the fusion module design is the primary determinant of cooperative perception quality, with any form of communication yielding ${\sim}58\%$ improvement over isolation; and (2)~R2T's deliberative approach provides the greatest benefit under high information asymmetry (heavy occlusion), where it approaches the oracle upper bound.
R2T takes a first step toward bridging deliberative reasoning and multi-agent cooperative perception, and we believe this direction holds significant potential as both fields continue to advance.

{
    \small
    \bibliographystyle{ieeenat_fullname}
    \bibliography{references}
}

\end{document}